\documentclass[authoryear, 2p, compress]{elsarticle}
\usepackage{graphicx} 
\usepackage{xcolor}
\usepackage{hyperref}
\usepackage{multirow}
\usepackage{array}
\usepackage{amsmath}
\usepackage{longtable}

\newcolumntype{C}[1]{>{\centering\arraybackslash}p{#1}}

\providecommand{\keywords}[1]{\textbf{Keywords:} #1}

\begin{document}

\begin{frontmatter}

\title{Predicting Long-Term Self-Rated Health in Small Areas Using Ordinal Regression and Microsimulation}

\author[nuig]{Seán Caulfield Curley\corref{cor1}}
\ead{s.caulfieldcurley1@universityofgalway.ie}
\cortext[cor1]{Corresponding Author}

\affiliation[nuig]{organization={School of Computer Science, University of Galway},
city={Galway},
country={Ireland}
}

\author[nuig]{Karl Mason}

\author[nuig]{Patrick Mannion}

\begin{abstract}
This paper presents an approach for predicting the self-rated health of individuals in a future population utilising the individuals' socio-economic characteristics. An open-source microsimulation is used to project Ireland's population into the future where each individual is defined by a number of demographic and socio-economic characteristics.
The model is disaggregated spatially at the Electoral Division level, allowing for analysis of results at that, or any broader geographical scales. Ordinal regression is utilised to predict an individual's self-rated health based on their socio-economic characteristics and this method is shown to match well to Ireland's 2022 distribution of health statuses. Due to differences in the health status distributions of the health microdata and the national data, an alignment technique is proposed to bring predictions closer to real values. It is illustrated for one potential future population that the effects of an ageing population may outweigh other improvements in socio-economic outcomes to disimprove Ireland's mean self-rated health slightly. Health modelling at this kind of granular scale could offer local authorities a chance to predict and combat health issues which may arise in their local populations in the future.
\end{abstract}

\end{frontmatter}


\section{Introduction}
\label{sec:introduction}

Accurate anticipation of the health trajectories of future populations is central to informing public health planning, healthcare resource allocation, and policy design. Self-rated health (SRH), a widely used measure that captures an individual’s overall perception of their health status, has been shown to predict morbidity \citep{chandola2000srhMorbidity}, healthcare utilization \citep{hansen2002srhUtilisation, chamberlain2014srhUtilisation2}, and mortality \citep{lorem2020srhAsAPredictor}. Its multidimensional nature, integrating both physical and psychosocial aspects of health, makes SRH a valuable outcome for population health forecasting.

Previous approaches using microsimulations to model health outcomes have largely either not focused on the temporal aspect of health (analysing the current population only) or relied on aggregate-level projections. In effect, the method proposed in this paper aims to combine these previous approaches to predict future health outcomes while incorporating the heterogeneity of health with respect to geography. 

\begin{figure}
    \centering
    \includegraphics[width=\linewidth]{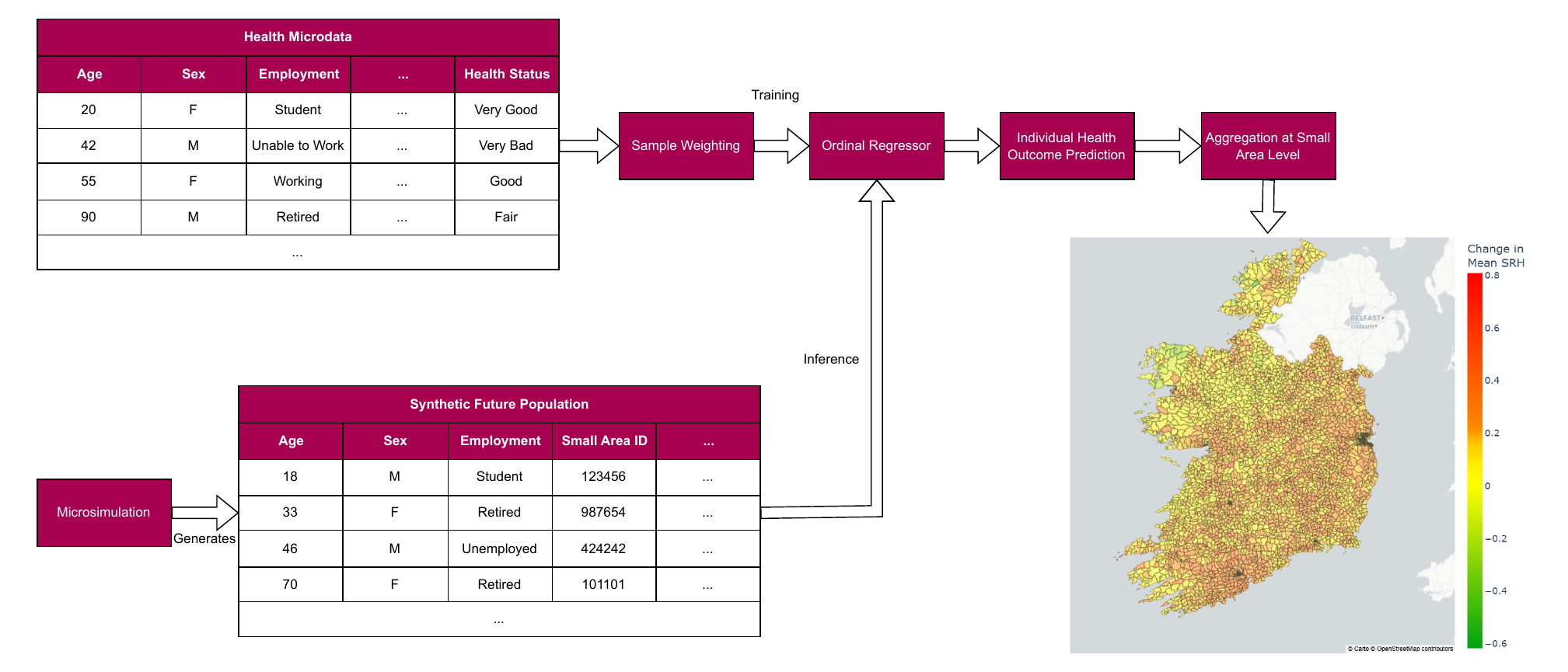}
    \caption{A graphical summary of how the approach operates. Note that synthetic individuals are created for this example and are not based on individuals from either dataset.}
    \label{fig:graphicalSummary}
\end{figure}

In this study, we integrate microsimulation with ordinal regression modelling to predict future distributions of self-rated health. Ordinal regression is well-suited for SRH, which is typically measured on an ordered categorical scale (e.g. ``Very Good'' to ``Very Bad''), allowing us to model the probability of individuals reporting different health levels based on their demographic profiles. By combining this regression framework with a microsimulation model, we may generate forward-looking estimates of population health under various demographic scenarios. A graphical summary of the approach is given in Figure \ref{fig:graphicalSummary}.

This work contributes to the field of health forecasting by providing a methodological approach that combines the strengths of microsimulation and ordinal regression. The resulting projections can support policymakers and public health professionals in anticipating future health needs, identifying vulnerable groups, and designing interventions that are responsive to demographic change. ``What-if'' scenarios can also be tested, both with regard to the microsimulation and the regression, meaning alternate scenarios than the ones proposed in this paper can be tested easily.

The rest of this paper is laid out as follows: Section \ref{sec:background} summarises the background to the problem, giving the motivation for using SRH, highlighting some relevant Irish microsimulations, and illustrating how microsimulations have been used in a health context previously. The next section goes in-depth in the workings of the approach. Each module of the microsimulation is summarised and the operation of ordinal regression is explained. The data used to support both methods is also described in detail. A method to align the health microdata with the observed Census data is introduced in subsections \ref{sec:aligning_per_cohort_srh_distributions} and \ref{sec:predicting_future_srh_distributions}. Section \ref{sec:results} presents the results for the approach, both in terms of validation and results of future projections. After this, a concise example application of the resulting populations is outlined in Section \ref{sec:case_study}. The paper concludes with a discussion on some of the strengths and limitations of the approach.

\section{Background}
\label{sec:background}

Microsimulation is a simulation paradigm in which simulation units represent low-level entities such as individual people, households, or farms. Dynamic microsimulations, where the temporal effect on the topic in question is analysed, are particularly powerful for analysing demographic changes at the individual level, and have been used to model population change in Canada \citep{demosimCanada2025}, Germany \citep{munnich2021Germany} and the US \citep{urbanInstitute2016dynasim}. One of the key outputs of microsimulation is enabling Agent-Based Modelling (ABM), where individuals in the population are treated as agents who can interact with each other and the environment. One example of an agent-based model being adapted from microsimulations is a model of urban freight distribution where microsimulation models passenger trips, and freight entities like customers, carriers and shippers are treated as agents \citep{gomez2018freight}. An example from demography is the model introduced by \citet{bae2016Korea}, which handles population size drivers such as fertility, mortality and migration with microsimulation and uses ABM to model the process of an individual finding a partner for marriage.

There have been multiple microsimulation studies published with Ireland as the focus. 
The Economic and Social Research Institute (ESRI) used survey and register data to model Ireland's tax-benefit system with a microsimulation \citep{keane2023esriTAX}. The Irish Farming Land Rental Market has also been modelled using the technique \citep{loughrey2022irishMicrosimLandRental}. \citet{odonoghue2013SMILE} pioneered microsimulation in Ireland by using a static model to create a synthetic population for the country. A dynamic, open-source model inspired by that model and those of the US, Canada and Germany was also recently released \citep{me2025irelandMicrosim}. A large contributing factor to the breadth of research performed to date in Ireland is the wealth of data provided by the Central Statistics Office (CSO). Data featured on the \url{data.cso.ie} repository includes Census data, Small Area Population Statistics (SAPS), and the results of the CSO's population projections. The CSO publish a report a few years after each Census outlining the results of their population projections \citep{cso2023projections} which are carried out using the Demographic Component Method (DCM). Their most recent report projects the Irish population from 2023 to 2057, disaggregated at the NUTS3 region level\footnote{The Nomenclature of Territorial Units for Statistics (NUTS) are a standard for referencing country subdivisions. In Ireland, the NUTS3 regions are Border, West, Mid-West, Mid-East, South-East, South-West, Dublin and Midlands}.

There have also been numerous applications of microsimulation to tackle health-related issues. \citet{wu2022GB} used static microsimulation techniques (without a temporal component) to create a synthetic population dataset relating health and socio-economic factors for small areas in Great Britain. A similar approach was used to generate a population which included both personal and national well-being as indicators of quality of life in Aotearoa New Zealand \citep{wiki2023qaleNZ}. \citet{campbell2016SimAlba} utilised the SimAlba model, which was developed for economic simulation in Scotland, to provide insights into geographically-sensitive health variables such as smoking, alcohol intake and mental well-being in small areas of Glasgow. The authors also demonstrate how such a model can be used to aid policy development by highlighting ``high-risk areas'' which would not have been possible previously. 

The reason for choosing SRH as the dependent variable in this study is that it has been found in  the literature to be a strong predictor of all-cause mortality \citep{lorem2020srhAsAPredictor, ganna2015srhAsAPredictor2}. With specific regard to the Irish context, a report published on \url{publicpolicy.ie} found that SRH is an accurate predictor of an individual's true health in terms of multimorbidity \citep{briody2015srhInIreland}. In terms of the predictive power of the independent variables in this paper (age, sex, marital status, economic status, education, region), a wide range of studies have shown strong predictive power (of SRH) in some or all of these characteristics across a spectrum of cultures \citep{jindrova2020srhAndSocioEconomicCharacteristicsSlovakia, kelleher2003srhAndSocioEconomicCharacteristicsIreland, prakash2025srhAndSocioEconomicCharacteristicsReview,darker2016srhAndSocioEconomicCharacteristicsDeprrivedSuburbanIreland,
kasenda2022srhAndSocioEconomicCharacteristicsMalawi}.

From, the literature, the study found to be closest to the approaches suggested in this paper uses a dynamic microsimulation of small areas to estimate future elderly health care demand \citep{clark2016estimateFutureElderlyHealthCareDemand}. However, there are a number of key differences between our work and that of \citep{clark2016estimateFutureElderlyHealthCareDemand}. Firstly, the microdata used by \citet{clark2016estimateFutureElderlyHealthCareDemand} is the English Longitudinal Study of Ageing and thus, all individuals in that study are aged 50 or older. Secondly, three specific morbidities were studied, namely cardiovascular disease, diabetes and respiratory illnesses. In our work, SRH is used to provide a more general picture of health for all individuals aged 15 or older. Finally, the projections of the structure of future populations in the paper by \citet{clark2016estimateFutureElderlyHealthCareDemand} focus on age, sex, and ethnicity only. Our work aims to extend that scope to include characteristics whose changes are more complex to model such as education and economic status. Inclusion of these additional variables also allows for leveraging of the most data possible and enables relatively simple interconnection with existing economic projections (such as those mentioned above). A similar paper to that by \citet{clark2016estimateFutureElderlyHealthCareDemand} also exists for older people in Ireland \citep{may2022projectingElderlyCareIrelandMicrosim}. Again, only adults aged 50 or above are considered and the paper focuses on the prevalence of specific diseases rather than general health. Education is included as an independent variable in that work, but no temporal changes in educational attainment are modelled in the microsimulation. Finally, the results are also not disaggregated geographically.

\section{Data and Methods}
\label{sec:data_and_methods}

\subsection{Microsimulation Populations}
\label{sec:microsimulation_populations}


\begin{table}[!htb]
    \centering
    \begin{tabular}{|C{0.475\linewidth}|C{0.475\linewidth}|}
    \hline
    \textbf{Characteristic} & \textbf{Possible Values}\\
    \hline
    Age & 0-105\\
    \hline
    Sex & Female, Male \\
    \hline
    Marital Status & Married, Single, Separated, Widowed \\
    \hline
    Citizenship & Ireland, UK, EU27 excluding Ireland, Rest of World\\
    \hline
    Moved to Ireland in last year & True, False\\
    \hline
    Highest Level of Education Attained & No Formal Education, Primary Education, Lower Secondary, Upper Secondary, Post-Leaving Certificate, Higher Certificate, Undergraduate Degree, Postgraduate Degree, Doctorate\\
    \hline
    Primary Economic Status & At Work, Student, Looking After Home/Family, Retired, Unable to Work Due to Permanent Sickness or Disability, Other, Unemployed, Not Applicable (for children)\\
    \hline
    \end{tabular}
    \caption{The characteristics used to describe individuals along with their possible values}
    \label{tab:characteristicsMain}
\end{table}

Geographically-diverse populations with all of the characteristics required for prediction of SRH are generated using the Socio-Economic Microsimulation for Irish population Projections (SEMIPro) model, which is described in detail elsewhere \citep{me2025irelandMicrosim}. The model assigns each individual the socio-economic characteristics age, sex, marital status, citizenship, whether the person moved to Ireland in the previous year, highest level of education attained and primary economic status. Mortality, internal and international migration, fertility, marriage, education and employment are all modelled. The CSO's DCM-based population projections were run until the year 2057, and so this is chosen as the end date for for SEMIPro also. Projections beyond 2057 would not have trustworthy results to validate against and the uncertainty of results also increases with each additional year. The possible values for each of the characteristics an individual possesses are given in Table \ref{tab:characteristicsMain}. A mapping between each characteristic value and its value in the source code is given in Appendix \ref{sec:characteristicValuesTable} for anyone who wishes to alter/expand the list of possible values.



\subsubsection{Assumptions}
\label{sec:assumptions}

The first step in the setup of the microsimulation is to choose which international migration scenario to simulate. These scenarios are the same as those from the CSO's Demographic Component Method population projections \citep{cso2023projections} and have the following values:
\begin{itemize}
    \item \textbf{M1}: Net migration starting at +75,000 and incrementally decreasing to +45,000 per annum by 2027. Net migration remaining at this level from 2027 on.
    \item \textbf{M2}: Net migration starting at +75,000 and incrementally decreasing to +30,000 per annum by 2032. Net migration remaining at this level from 2032 on.
    \item \textbf{M3}: Net migration starting at +75,000 and incrementally decreasing to +25,000 per annum by 2027 and +10,000 by 2032. Net migration remaining at this level from 2032 on.
\end{itemize}
As in the CSO's regional population projections \citep{cso2023projections}, 1 internal migration scenario is modelled. The county-to-county migration flows for this scenario are calculated from the averages of the 2016 and 2022 Census values of county's populations categorised by usual residence one year prior.

\subsubsection{Initialisation}
\label{sec:initialisation}

Once the simulation scenarios have been chosen, initialisation of the microsimulation begins. Initially, a representative static population for 2022 is retrieved from the Irish synthetic population datasets \citep{me2025syntheticPop}. The population generated using Iterative Proportional Fitting (IPF) \citep{deming1940IPF} is utilised here, as it is the method which achieved the best goodness-of-fit on the Census aggregate data at the Electoral Division level. The first step in initialisation of the population is to assign a spouse to each individual with a marital status of ``Married''. Then, the level of education that individuals with the primary economic status of ``Student'' are studying for is estimated. It is then assumed that the student has achieved the highest level of education up to that point e.g. a person studying for an Upper Secondary qualification is assumed to have achieved a Lower Secondary qualification. The student is also assigned a graduation date, determined from the average completion time for each level of education.

\subsubsection{Mortality}
\label{sec:mortality}

Next, the dynamic portion of the microsimulation begins. The ordering of modules in the microsimulation is chosen in such a way that maximises the information available to each module. For example, the transition probabilities for a person’s economic status are dependent on their age, sex, citizenship and education level. Therefore, the employment module is computed after every other module to allow for changes in age and education level in the same year. Mortality and ageing is the first module simulated. In this module, individuals are assigned a probability of dying based on their age, sex, and the current year. The current year is included to follow the expected improvements in mortality anticipated in the CSO's population projections. All survivors' ages are then incremented by one year.

\subsubsection{Migration}
\label{sec:migration}

The next module simulated is internal migration. In this module, intra- and inter- county migrants are moved according to the migration flows observed in the year leading up to the most recent Census. Migrants' ages and sexes are sampled proportionally to the observed distributions in the 2022 Census. International migration is the next module to be simulated. The total number of international emigrants and immigrants is controlled by the international migration scenario decided on at the beginning of the microsimulation. The first step in this module is to split Ireland's population into its NUTS3 region subpopulations to model the correct proportions of national emigrants from each region. The percentage of a region's population emigrating is calculated according to the average numbers between 2017 and 2022. Emigrants from each region are then sampled proportionally to their age and sex and these emigrants are removed from the population. 

As is the case for international emigrants, the age and sex of international immigrants is sampled according to the proportions of those characteristics observed in international immigrants to Ireland between 2017 and 2022 (inclusive). The 'Moved to Ireland in the Last Year' characteristic is used to accurately sample international immigrants. Accurate Sampling is achieved by using an individual with the same age, sex, and citizenship from this recent immigrant population as a donor for the individual's socioeconomic characteristics. The destination ED of international immigrants is randomly sampled according to each ED’s proportion of the total number of people whose usual residence was outside of Ireland one year before the census.

\subsubsection{Fertility and Marriage}
\label{sec:fertilityAndMarriage}

Fertility is then modelled by following the CSO's assumption of a decrease in the national Total Fertility Rate (TFR) from 1.55 in 2022 to 1.3 by 2038. The TFR is assumed to remain constant after 2038. Fertility rates are disaggregated by age, sex, NUTS3 region, and marital status according to 2022 values. Each newborn is assigned an age of 0, a randomly chosen sex, a marital status of ``Single'', and both an economic status and education of ``NA''. The baby is assigned to the same ED’s population as their mother who is randomly sampled from the relevant subpopulation. A citizenship value of ``IE'' is given to the newborn and their ``resided outside Ireland one year ago'' characteristic is set to False. Marriages and separations are then simulated. The number of married couples to be separated is calculated based on the total number of married people and the total number of separations in 2022. Unmarried individuals are then matched into married couples based on their NUTS3 region, age, sex, and highest level of education attained.

\subsubsection{Education and Employment}
\label{sec:educationAndEmployment}


\begin{figure}[!htb]
    \centering
    \includegraphics[width=\linewidth]{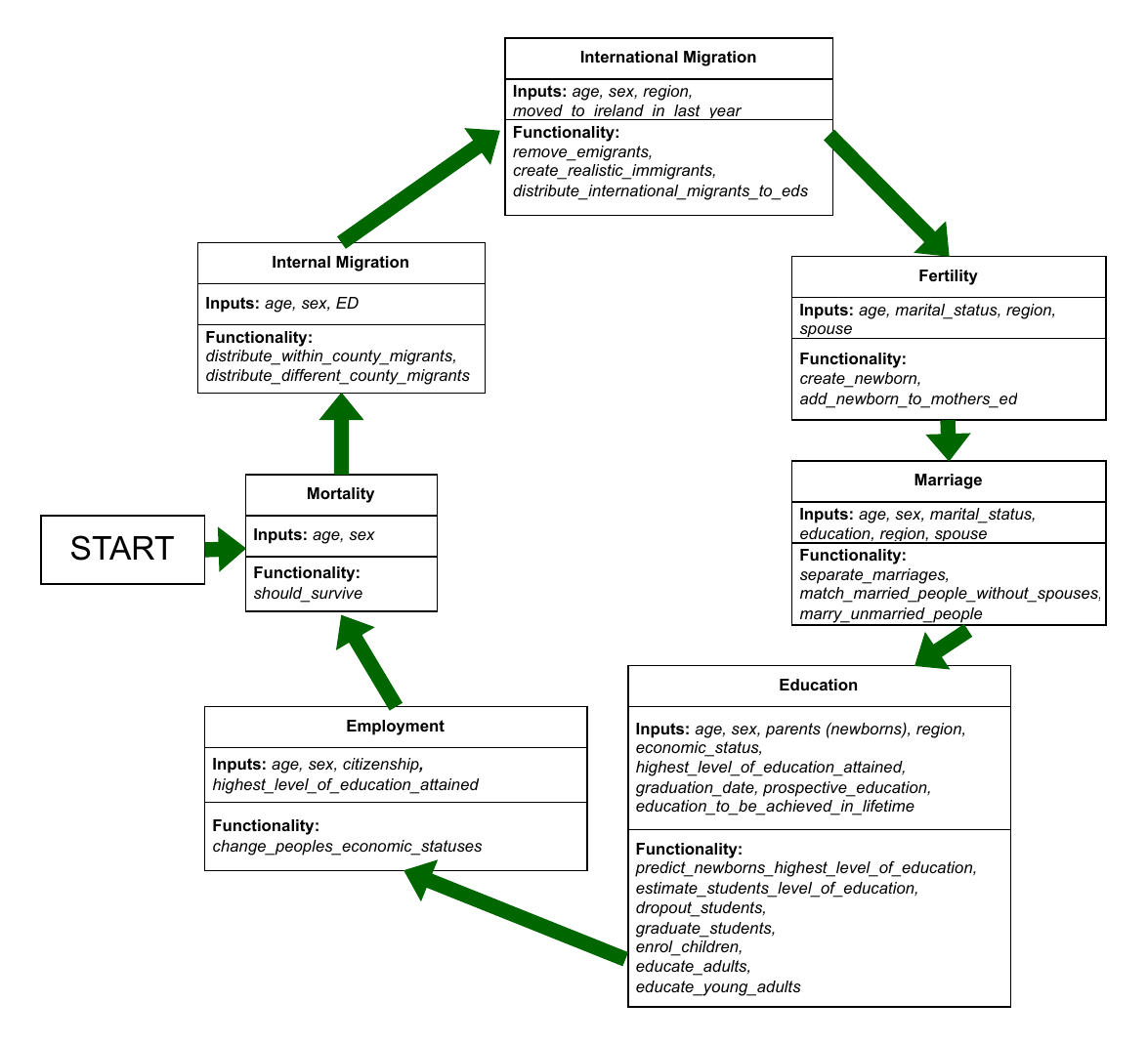}
    \caption{A flowchart summarising the workings of the SEMIPro Irish microsimulation model \citep{me2025irelandMicrosim}.}
    \label{fig:MicrosimFlowchart}
\end{figure}

Education is a complex module in the microsimulation owing to the large amount of data available. Firstly, the highest level of education a newborn will achieve in their lifetime is estimated based on the highest level of education achieved by their parent(s). 
Next, some students are selected to drop out of their courses. The number of students to drop out from each level of education is decided according to the observed dropout rates from 2022. 
Selection of which students should drop out is performed by prioritising those with a ``highest level of education to be attained in their lifetime'' equal to or less than the level of education they are currently studying for. 
The graduation dates of all remaining students are then checked, and the highest level of education attained for students projected to graduate in the current year is updated to their prospective qualification. The next step, possible changing of economic status, for both dropouts and graduates is then simulated, based on the level of education the student was studying for and whether they dropped out or graduated.

The final two steps in the education module involve adding children of school enrolment age (4 years old) into Primary School and adding a number of adults back into the education system. The proportion of adults between the ages of 25 and 69 that were in formal education in 2022 is used to calculate the number of older students to be added in each NUTS3 region. Which adults to add back into education is stratified based on age and sex, and the level of education an adult learner will study for is based on their current level of education. Finally, the economic statuses of all non-student adults are transitioned based on the individual's age, sex, citizenship, and education. A graphical summary of the microsimulation is given in Figure \ref{fig:MicrosimFlowchart}.
 
\subsection{Ordinal Regression}
\label{sec:ordinal_regression}
Ordinal regression is a form of regression performed when the dependent variable can be ranked e.g. ``star'' ratings for movies, levels of agreement with a statement (Likert scale), etc \citep{tutz2022OrdinalRegressionReview, burkner2019OrdinalRegression}. If one was to treat the dependent variable as a label for classification, information would be lost. For example, in this experiment the dependent variable (SRH) can have the values ``Very Good'', ``Good'', ``Fair'', ``Bad'' or ``Very Bad''. A model which predicted an SRH of ``Bad'' when the true label was ``Good'' should probably be punished more than one which predicts an SRH of ``Very Good''. However, a classification model would treat both the ``Very Good'' and ``Bad'' prediction as simply ``wrong''. Conversely, a pure regression model where the SRHs would be coded as integers is also not appropriate. This is because there is no intuition behind any of the real numbers between integers which the regression model would predict (What does $1.33$ represent if ``Good'' is 1 and ``Fair'' is 2?). Another issue is that responses may not be equidistant. For example, an individual rating a movie may feel there is a much larger gap between rating a movie with 5 stars instead of 4 than there is with rating a movie with 4 stars instead of 3.

There are three main types of ordinal regression model: cumulative models, sequential models and adjacent category models. Cumulative models are the most widely used form of model and are useful for data such as the aforementioned Likert scale. Sequential models, on the other hand, are employed for tasks where reaching an ordered category, $k$, implies that all categories less than $k$ have also been reached. For example, in a model predicting the distance a car can travel before running out of fuel, we know that to drive $k$ kilometres away, one necessarily had to $k-1$ kilometres, $k-2$ kilometres, etc. The final type of model, adjacent category models, are less naturally inspired and therefore intuitive examples of real-world applications are difficult to describe. They are usually employed when proportional odds assumptions do not hold. For example, one proportional odds assumption is that predictors have the same effect on all response categories. However, in the example of a person rating a movie on a 5-star scale, age may not have much influence on whether the person rates the movie 1 or 2 stars. However, age may have a large influence on whether they rate the movie with 4 or 5 stars.

In this experiment, a cumulative model will be employed and thus, we assume that the observed ordinal variable (people's SRH) is underlain by a latent (non-observable) continuous variable $Y^{*}$. Then, we assume that $Y^{*}$ can be partitioned into the $K+1$ categories of $Y$ by latent thresholds, denoted $\tau_k$ $(1\leq k \leq K)$. Then, $Y$ is equal to category $k$, if and only if $Y^{*}$ is greater than $\tau_{k-1}$ and less than or equal to $\tau_{k}$. The thresholds of $Y^{*}$ are governed by the equation $-\infty = \tau_0 <\tau_1<\ldots<\tau_k<\tau_{k+1}=\infty$. As mentioned above, we cannot assume the categories of $Y$ to be spaced equally within the respondent's mind. However, we can assume this with $Y^{*}$, and therefore regression can be performed on it. The equation used to represent the regression model is

\begin{equation}
    Y^{*}=-\textbf{x}^T \boldsymbol{\beta} - \epsilon
\end{equation}

where $\textbf{x}$ is the vector of explanatory variables, $\boldsymbol{\beta}$ is a vector of coefficients and $\epsilon$ is a noise variable with continuous distribution function $F(.)$. In probit models, $\epsilon$ is a standard normal distribution while for logit models, $\epsilon$ is a standard logistic distribution.

All of the above leads to the cumulative model

\begin{equation}
    P(Y \leq k|\textbf{x})=F(\beta_{0k} + \textbf{x}^T\boldsymbol{\beta}), \quad k=1,\ldots, K
\end{equation}

where $\beta_{0k}$ are the category-specific intercepts, and are equal to the thresholds on the latent scale, $\beta_{0k}=\tau_k$. As mentioned, logit models use the logistic distribution $F(\eta) = exp(\eta)/(1+exp(\eta))$ which gives us

\begin{equation}
    logitP(Y \leq k|\textbf{x})=\beta_{0k} + \textbf{x}^T\boldsymbol{\beta}
\end{equation}

We are interested in finding the probability of the response being a single category, $k$. This can be calculated simply by subtracting the cumulative probability of $Y$ being less than or equal to $k$ from the probability of $Y$ being less than or equal to $k+1$.



\begin{figure}[!htb]
    \centering
    \includegraphics[width=0.6\linewidth]{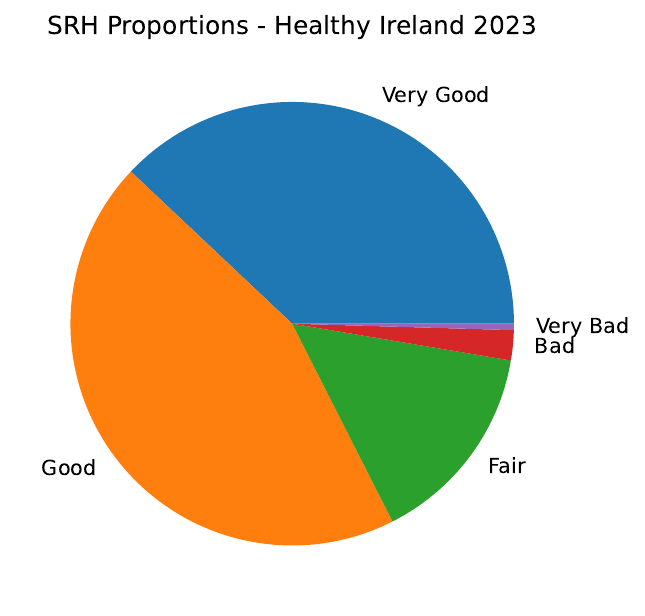}
    \caption{The overall distribution of SRHs from the Healthy Ireland 2023 survey}
    \label{fig:srhPieChart}
\end{figure}

The Healthy Ireland Survey 2023 \citep{deptOfHealth2025HealthyIreland} acted as the source for the microdata needed for ordinal regression. The survey is commissioned by the Department of Health and is carried out annually. There were approximately 7,500 respondents to the survey in 2023, all of whom were aged 15 or older. Respondents answer questions covering a range of subjects including general health, smoking, health service utilisation and more. A number of demographic characteristics for each respondent were also tracked. As the microdata is highly sensitive, only researchers approved by the Irish Social Science Data Archive may access the records. In this study, the survey's first question ``How is your health in general?'' is in focus. Respondents were given the options ``Very Good'', ``Good'', ``Fair'', ``Bad'', ``Very Bad'', ``Don't know'' or refusal to answer. The small percentage of respondents who did not answer or answered ``Don't know'' were excluded from this experiment. The distribution of SRHs from the survey is given in Figure \ref{fig:srhPieChart}.

The independent input variables to the ordinal regressor are the individuals' age group, sex, marital status, primary economic status, highest level of education attained and region of residence. Citizenship and 'resided outside Ireland one year ago' are not used owing to a lack of data in the microdata. The Healthy Ireland survey does include data on individuals' country of birth, but because of the possibility of individuals possessing citizenship to countries other than their birthplace, it was decided not to treat these characteristics as equivalent. The regions given in the microdata are ``Connacht/Ulster'', ``Munster'', ``Dublin'' and ``Leinster Rest''. Prediction of the health status of an individual is achieved by predicting the individual's probabilities of having each of the possible SRHs given their value for each of the mentioned characteristics, and then sampling probabilistically from this output. 

For ease of compatibility with the microsimulation data, the Python module \verb|statsmodels| is utilised for the ordinal regression in this study. However, one could easily implement the same approach using the \verb|ordinal| or \verb|MASS| packages in R or the \verb|ologit| command in Stata. 

\subsection{Aligning Per-Cohort SRH Distributions}
\label{sec:aligning_per_cohort_srh_distributions}
One issue that arose in the development of this model was a mismatch between the SRH distributions in the microdata and the national proportions from the Census. In order to address this, each prediction was altered by the ratio of the national data to the microdata. For example, consider a scenario where the distribution of health statuses for a given cohort in the microdata was $[0.3, 0.4, 0.2, 0.08, 0.02]$ and the distribution for the same cohort in the Census was $[0.5, 0.3, 0.1, 0.05, 0.05]$. A prediction of $[0.45, 0.35, 0.1, 0.05, 0.05]$ would be adjusted by dividing the census distribution by the microdata distribution element-wise, and then multiplying each ratio by its corresponding predicted proportion. The result is then re-normalised to sum to 1. In the example, the final (adjusted) prediction would be approximately $[0.615, 0.215, 0.041, 0.026,$
$0.103]$.

\subsection{Predicting Future SRH Distributions}
\label{sec:predicting_future_srh_distributions}

\begin{table}[!htb]
    \centering
    \begin{tabular}{|C{0.45\linewidth}|C{0.45\linewidth}|}
    \hline
    \textbf{Historical Data} & \textbf{Microdata} \\ \hline
        Unemployed looking for first regular job & \multirow{2}{*}{Unemployed} \\
        Unemployed having lost or given up previous job & \\ \hline
        Student or pupil & Student or pupil \\ \hline
        Looking after home/family & Looking after home/family \\ \hline
        Retired & Retired \\ \hline
        Unable to work due to permanent sickness or disability & Unable to work due to permanent sickness or disability \\ \hline
        Employer or own account worker & \multirow{3}{*}{Working} \\ 
        Employee & \\
        Assisting relative & \\ \hline
        Other & Other \\ \hline
    \end{tabular}
    \caption{Economic statuses in the historical data and their counterparts in the microdata.}
    \label{tab:historic_to_microdata_economic_statuses}
\end{table}

The future distribution of health statuses is unknown, inherently. Therefore, a prediction model must be trained to estimate potential distributions of SRH for each cohort. Data from the previous three Censuses \footnote{Per-cohort SRH distribution data taken from \url{https://data.cso.ie/table/F4024}} are used as the basis for prediction of future SRH distributions. In this case, cohorts refer to the usually resident population of Ireland at least 15 years old separated by their 5-year age group, sex and principal economic status. The cohorts used in the prediction process do not group individuals by all of the possible characteristics. This means that SRH variations amongst individuals sharing an age group, sex and economic status but with differing marital status, education or region will be captured. This allows for increased diversity in the predictions of future individuals' SRHs. As in Section \ref{sec:microsimulation_populations}, some pre-processing is needed to transform the historical data's economic statuses into the form given in the health microdata. The transformations from the historical data to the health microdata are summarised in Table \ref{tab:historic_to_microdata_economic_statuses}.

\begin{equation} 
\label{eq:alr}
    alr(x) = \left[ln\frac{x_1}{x_D}, ...,ln\frac{x_{D-1}}{x_D} \right]
\end{equation}

\begin{equation}
    \label{eq:inverse_alr}
    alr^{-1}(y) = C\left[exp([y_1,y_2,...,y_{D-1},0])\right]
\end{equation}

\begin{equation}
    \label{eq:closure}
    C[x]=\left[\frac{x_1}{\sum^{D}_{i=1}x_i},...,\frac{x_D}{\sum^{D}_{i=1}x_i}\right]
\end{equation}

The SRH distributions are an example of compositional data (proportions of a whole) and have a constraint that the proportions for a cohort must sum to $1.0$. However, incorporating this constraint into time-series models is a difficult problem. Therefore instead, the proportions are transformed using the additive log ratio (ALR) transformation. Equation \ref{eq:alr} shows how each proportion $x_i$ is transformed, where $x_i$ denotes the number of people fitting category $i$ divided by the total number of people and where $D$ is the number of proportions in the compositional data. Any proportion can be used as the denominator $x_D$, and for this experiment the final proportion, Very Bad SRH, was chosen. Using ALR then allows us to predict the proportions as unbounded real values, which can be transformed back to proportions using the inverse transformation given in Equations \ref{eq:inverse_alr} and \ref{eq:closure}.

Prediction of the future log-ratios is done using a separate Gaussian Process (GP) for each dimension and for each cohort. The models take the Census year and the corresponding SRH distribution for that cohort for that year as input. GP regressors are used here because they give us an estimate of uncertainty along with the mean predicted value. Thus, we can use Monte Carlo simulation to sample a range of possible values for the future distributions of SRH. Here, 1000 samples per cohort are used. We report on the mean prediction as well as possible best and worst case scenarios. Best- and worst-case scenarios are calculated using a 90\% confidence interval, where the best-case would represent the predicted distribution with a ``Very Good'' proportion in the 95th percentile. Then, the distribution with a ``Very Good'' proportion in the 5th percentile would be assigned as the worst-case scenario.

\subsection{Experiments}
\label{sec:experiments}
The preceding section has summarised the operation of the SEMIPro microsimulation model and outlined how ordinal regression models can be used to predict ordinal characteristics like SRH. The need for alignment of SRH distributions to the observed national values as well as a method for projecting these SRH distributions has also been presented. In the next section, the approach is validated by comparing each ED's predicted mean SRH in 2022 to its actual mean SRH in 2022. The following section outlines an analysis of which characteristics are most influential in the prediction of SRH, both detrimentally and positively. 

Section \ref{sec:future_srh_distributions_results} presents results from the ALR GP process outlined in Section \ref{sec:predicting_future_srh_distributions}. From these results, an average, worst and best-case scenario are identified. Then, the overall SRH outlook for each scenario is investigated at the national level. In Section \ref{sec:small_area_results}, the fine-grained nature of the approach is leveraged to analyse results for individual EDs, as well as trends at the ED level. Finally, a case study is presented in Section \ref{sec:case_study}. Here, the mean SRHs of EDs are compared with the ED's distance from the nearest adult emergency department. The intention of this case study is to highlight the usefulness of the geographical diversity of the model's results. It is found that there are neighbouring EDs with notably different SRH outlooks, and the considerations about health facility planning these results may motivate are discussed. 

\section{Results}
\label{sec:results}

Results are presented for the 2057 population generated in the M1 International Migration and 2022 Internal Migration scenario. The described approach would work for any microsimulation scenario. Note that where numerical values are used for SRH, 0 corresponds to ``Very Good'' SRH while 4 corresponds to ``Very Bad'' SRH. Therefore, an improvement in SRH would be represented as a decrease in the numerical SRH value, and a disimprovement in SRH would be represented as an increase in the numerical SRH value. As in the results presented for SEMIPro itself \citep{me2025irelandMicrosim}, the average results from 5 runs of the microsimulation for the same scenario are presented in this section. 

\subsection{Validation}
\label{sec:validation}


\begin{figure}[!htb]
    \centering
    \includegraphics[width=\linewidth]{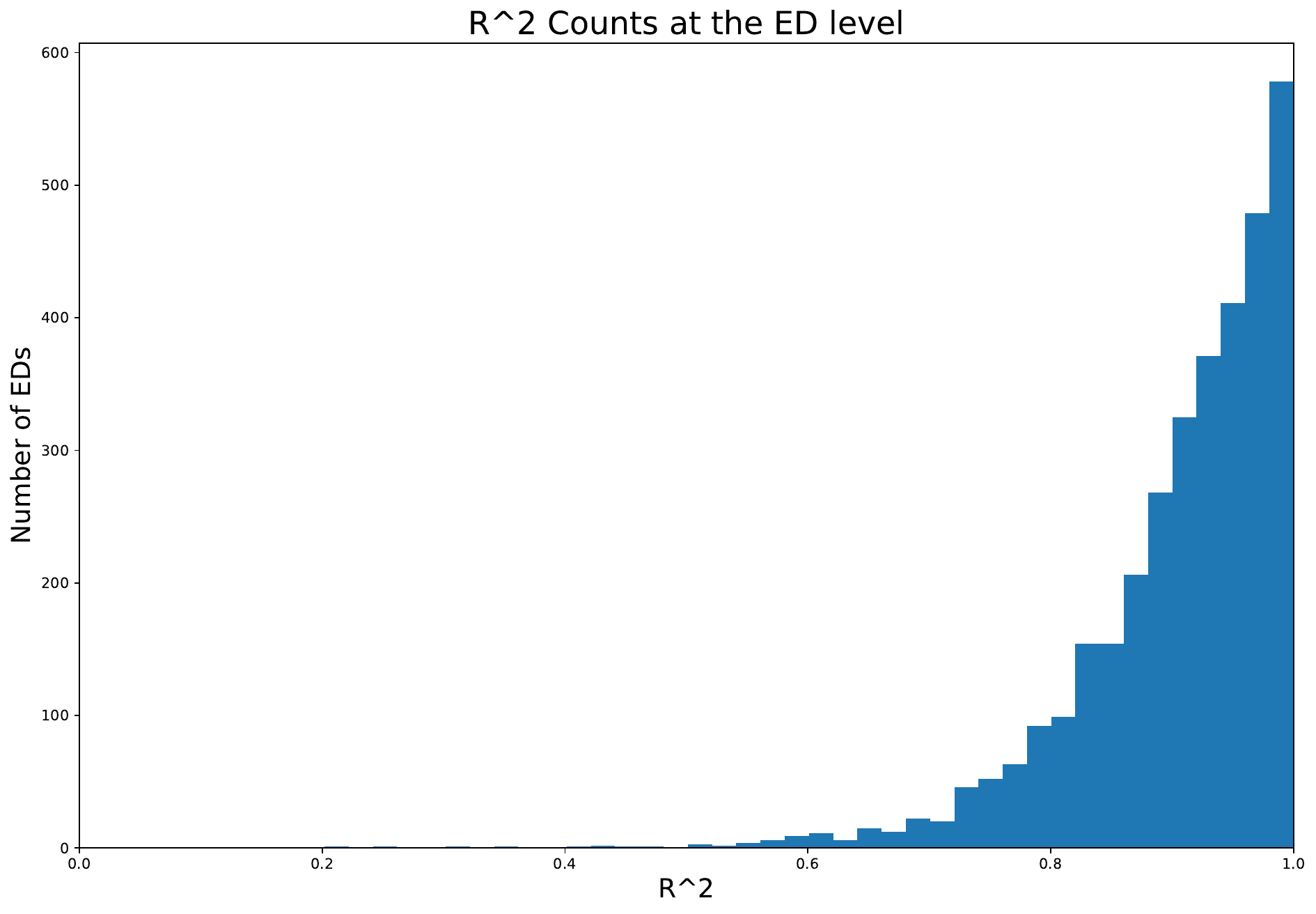}
    \caption{Distribution of $R^2$ values across the 3,417 EDs}
    \label{fig:r2_histogram}
\end{figure}

Validation of the approach is performed by comparing each ED's predicted distribution of SRHs to its actual distribution from the 2022 Census. The $R^2$ and Mean Squared Error (MSE) for each ED are then then calculated. The mean $R^2$ is determined to be approximately $0.9$ while the mean MSE is approximately $0.0037$. Figure \ref{fig:r2_histogram} reveals that the distribution of $R^2$ values across all EDs is heavily skewed towards a value of $1.0$. This high $R^2$ value and low MSE indicates that our approach is accurate at estimating an area's overall SRH based on actual data.

\subsection{Feature Importances}
\label{sec:feature_importances}


\begin{figure}[!htb]
    \centering
    \includegraphics[width=\linewidth]{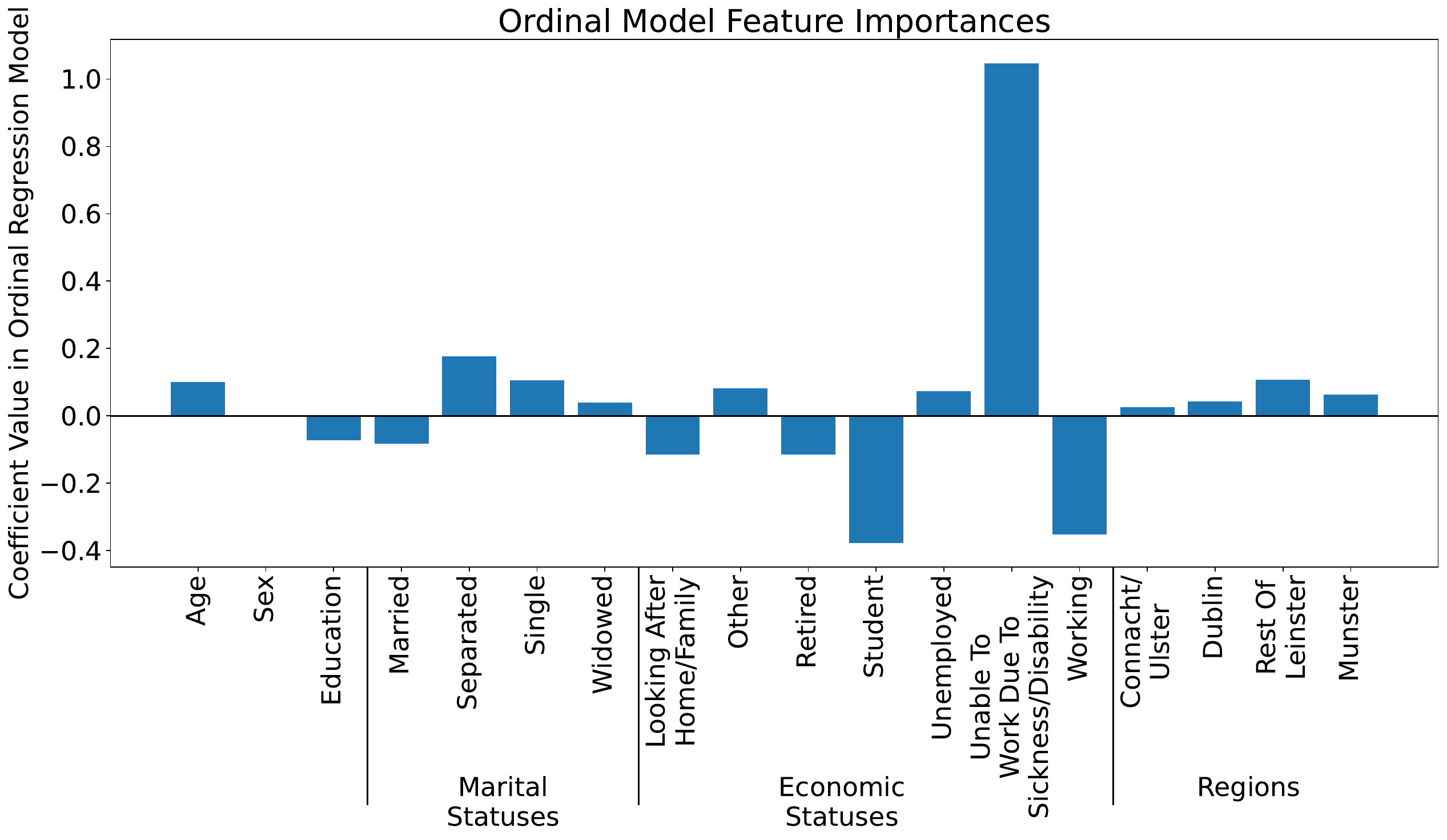}
    \caption{The effect of each characteristic on the predicted SRH. Characteristics which cannot be assigned a numerical or boolean value are labelled with their overall category.}
    \label{fig:featureImportance}
\end{figure}

Next, we analysed the importance of different input characteristics for the prediction of SRH.
One benefit of using \verb|statsmodels| is the possibility of generating a summary of the fitted model's parameters using the \verb|summary()| function. This function allows the user to examine the predictive power of each input characteristic, along with whether the characteristic has an increasing or decreasing effect on the target. The results of this analysis are presented in Figure \ref{fig:featureImportance}.
The primary economic status ``Unable to work due to sickness or disability'' is projected to be by far the strongest predictor of disimproving SRH, which is an intuitive result. This is followed by the marital status of ``Separated'', residing in the rest of Leinster, and the marital status of ``Single''. The three strongest positive predictors of SRH are the economic statuses of ``Student'', ``Working'' and ``Looking after Home or Family''. 

\begin{figure}[!htb]
    \centering
    \includegraphics[width=0.7\linewidth]{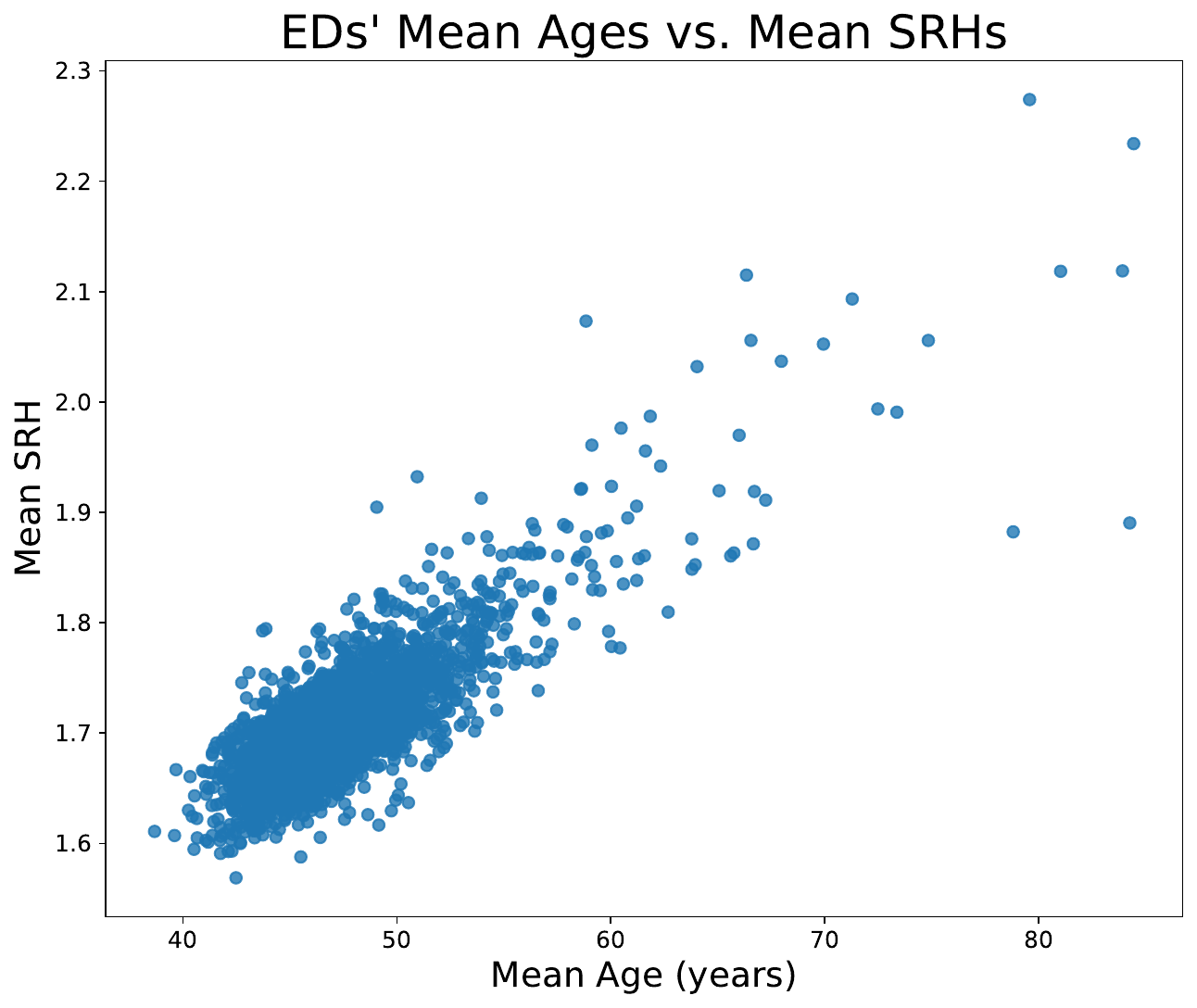}
    \caption{Each ED's mean age compared to their average SRH in the mean scenario.}
    \label{fig:ageVsSRH}
\end{figure}

\begin{figure}[!htb]
    \centering
    \includegraphics[width=0.7\linewidth]{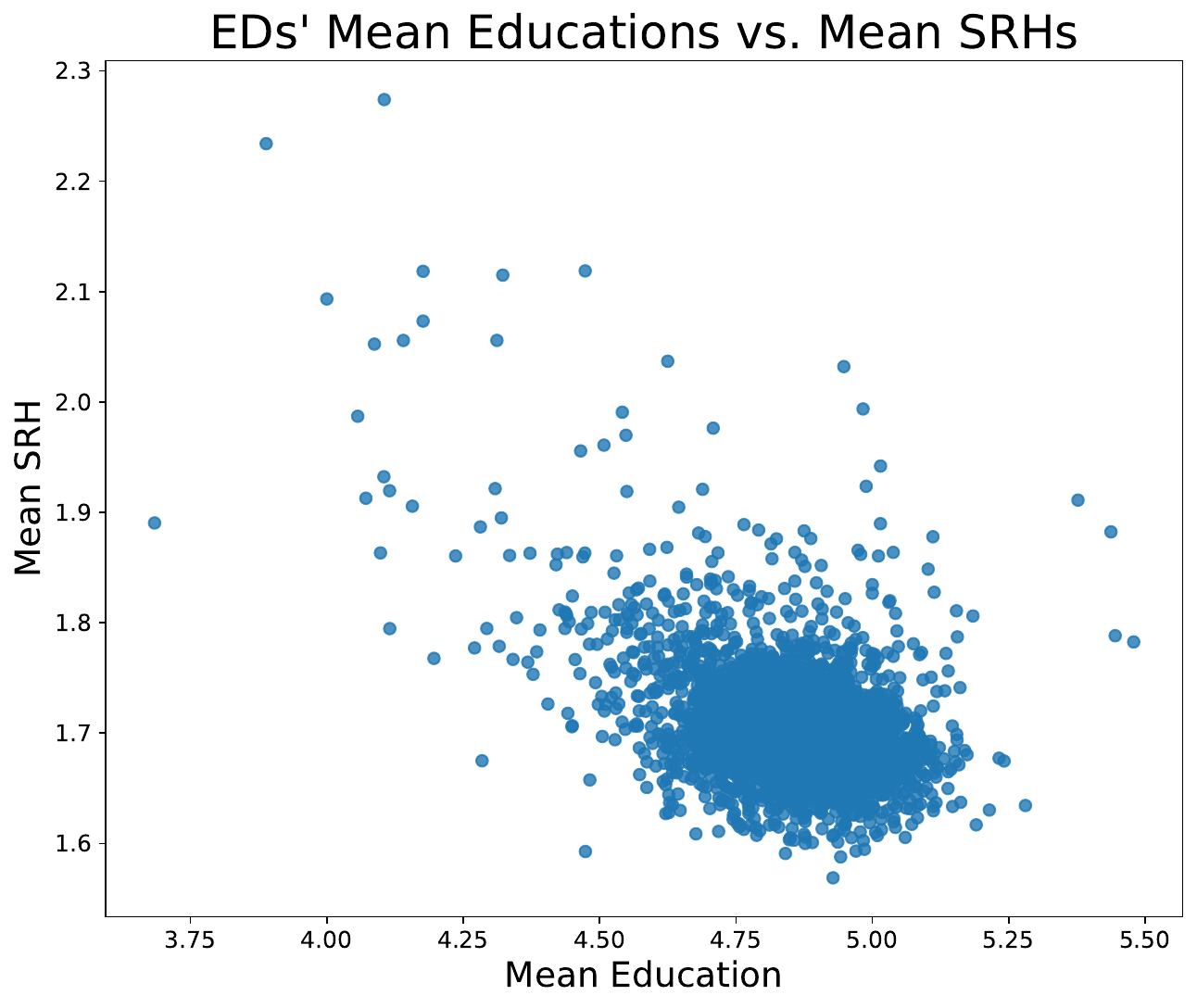}
    \caption{Each ED's mean education compared to their average SRH in the mean scenario. Here, education is represented on a scale where 0 represents no formal education and a doctorate is represented by 8.}
    \label{fig:educationVsSRH}
\end{figure}

In order to investigate the effects of these predictions on ED's mean values and to further verify the correctness of the model, each ED's mean age and education was compared to their mean SRH in the mean scenario. Figure \ref{fig:ageVsSRH} illustrates the expected result that increasing age is a strong negative predictor of SRH \citep{andersen2007ageAndSRH, ZAJACOVA2017ageAndSRH2}. In Figure \ref{fig:educationVsSRH}, the mean education of each ED is plotted against its mean SRH. An overall trend towards improving education being associated with better SRH can be observed. This is another encouraging result, given that improving education has been shown to be a strong positive predictor of improving SRH in the previous literature \citep{ross1995educationAndSRH}.

\subsection{Future SRH Distributions}
\label{sec:future_srh_distributions_results}

\begin{figure}[!htb]
    \centering
    \includegraphics[width=\linewidth]{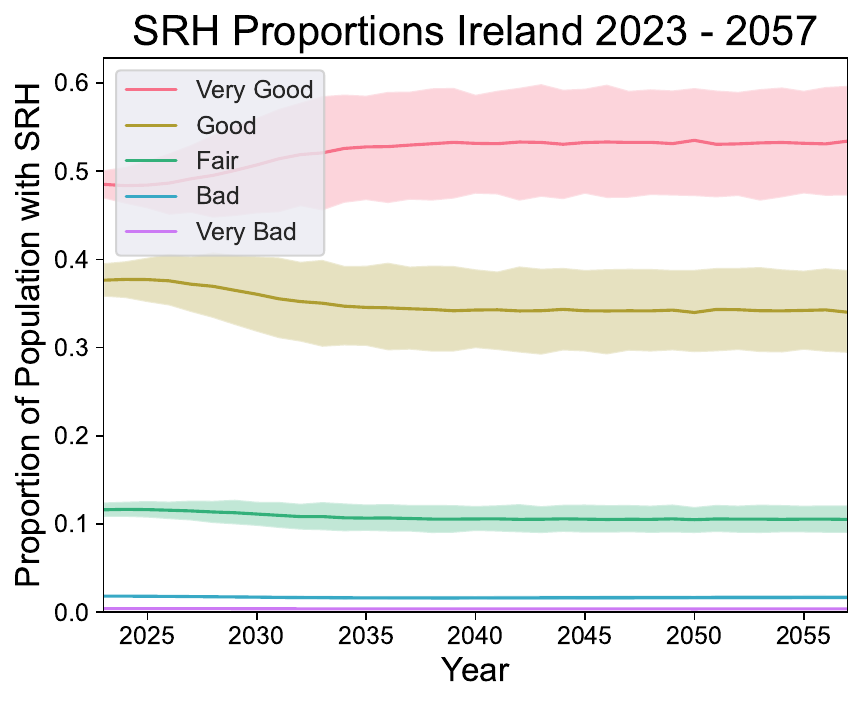}
    \caption{Predicted changes in the overall distribution of SRHs in Ireland up to 2057. Shaded areas represent 90\% confidence intervals for the given SRH status}
    \label{fig:national_proportions_prediction_with_uncertainties}
\end{figure}

When calculating the alignment values for each cohort of the future population, the range of SRH distributions is calculated as outlined in Section \ref{sec:predicting_future_srh_distributions}. However, in the interests of concision, results are not presented for each of the 84 possible cohorts. Instead, the same methodology is applied to the national distribution of SRHs as an indicator of the trends at the cohort level. Figure \ref{fig:national_proportions_prediction_with_uncertainties} demonstrates how each SRH status is predicted to evolve from 2023 to 2057. Higher uncertainty (a larger shaded area surrounding the line) for the two most prevalent SRH statuses (very good and good health) is intuitive, given that a change of, e.g., 20\% in one of these values will correspond to a much larger change in overall proportion than in one of the less prevalent SRH statuses. 

With regard to the changes themselves, we see a gradual increase in the proportion of the population rating their health as Very Good and a decline in those rating their health as Good until approximately 2035, at which point both proportions level off. This is very likely to represent the GP returning to the mean of the respective health status from the training data, and so it in itself does not provide particularly interesting insights. For example, the proportion of people reporting their health as Very Good was slightly lower in the 2022 Census than those in the 2011 or 2016 Censuses so the model gradually climbs back to the mean of the 3 values. However, calculating the model's uncertainty is useful as it gives us an indication of the spectrum of possible future SRH distributions, including allowing us to suggest best- and worst-case scenarios. Furthermore, Ireland's SRH distribution has been relatively stable since 2011\footnote{See \url{https://data.cso.ie/table/F4024}}, so continued stability is not out of the realm of possibility.

\subsection{National-Level Results}
\label{sec:national_level_results}


\begin{figure}[!htb]
    \centering
    \includegraphics[width=\linewidth]{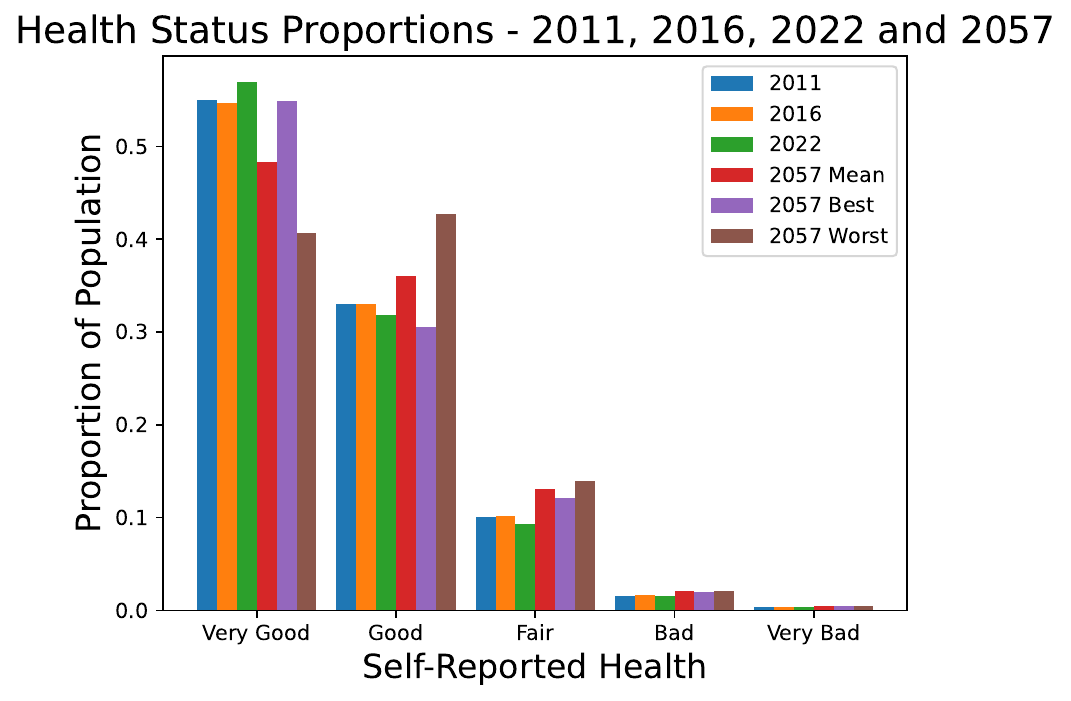}
    \caption{Overall SRH proportions for the past three Censuses and the predicted proportions for 2057}
    \label{fig:proportions_2011_2016_2022_2057}
\end{figure}

As mentioned in Section \ref{sec:future_srh_distributions_results}, the distribution of SRH statuses in Ireland has been relatively stable across the last three Censuses. This trend is observed in Figure \ref{fig:proportions_2011_2016_2022_2057}. However, on average, a relatively large change in these proportions is predicted for 2057. The mean projected national SRH proportions see a significant rise in the number of people reporting their health as Good or Fair with a sizeable decline in the share of people rating their health as Very Good. In fact, even in the best-case scenario, this share of the population is not projected to reach the same value as in any of the previous three censuses. In the worst-case scenario, it is projected that more people will report a health status of Good than Very Good (for the first time in the Irish Census). 

\subsection{Small Area Results}
\label{sec:small_area_results}



\begin{figure}
    \centering
    \includegraphics[width=0.95\linewidth]{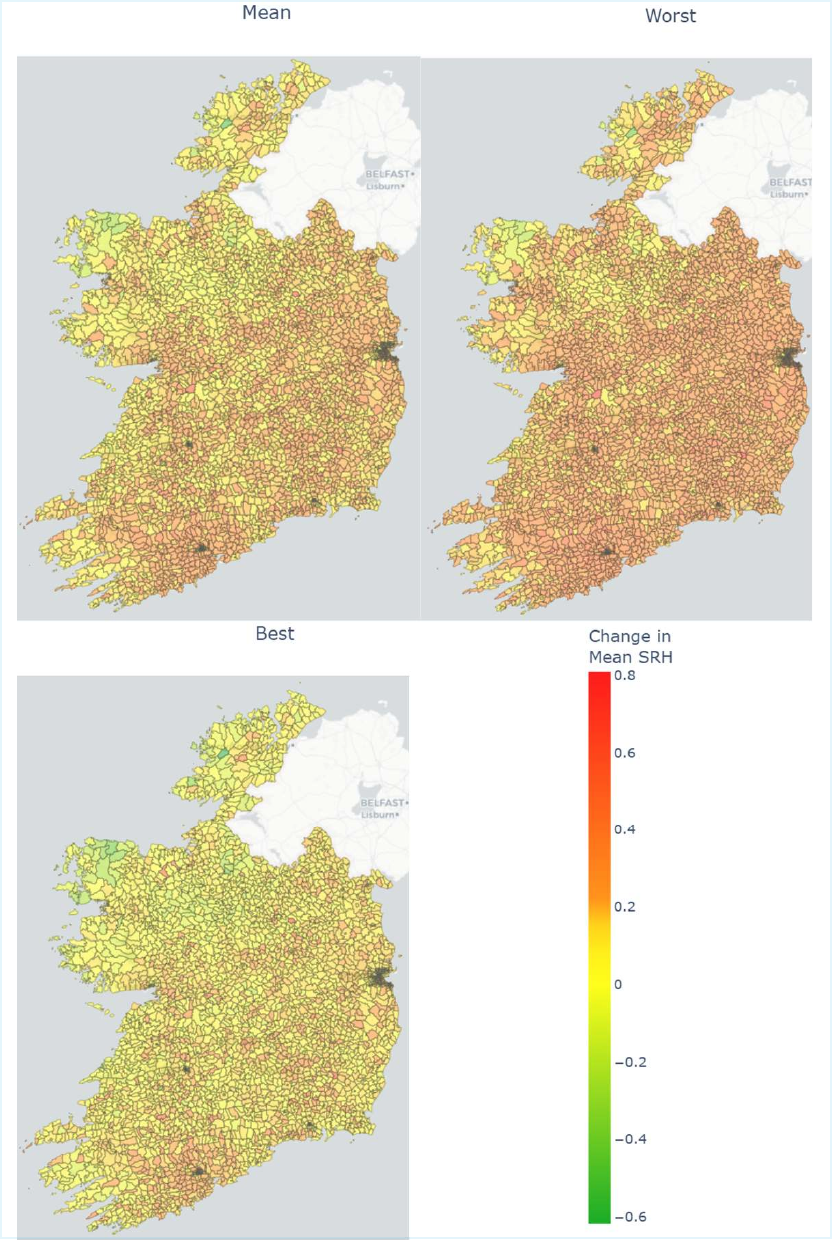}
    \caption{Changes in mean SRH per ED for the mean, best and worst case scenarios. All three scenarios are plotted on the same colour scale.}
    \label{fig:sideBySideChoropleths}
\end{figure}


\begin{figure}[!htb]
    \centering
    \includegraphics[width=\linewidth]{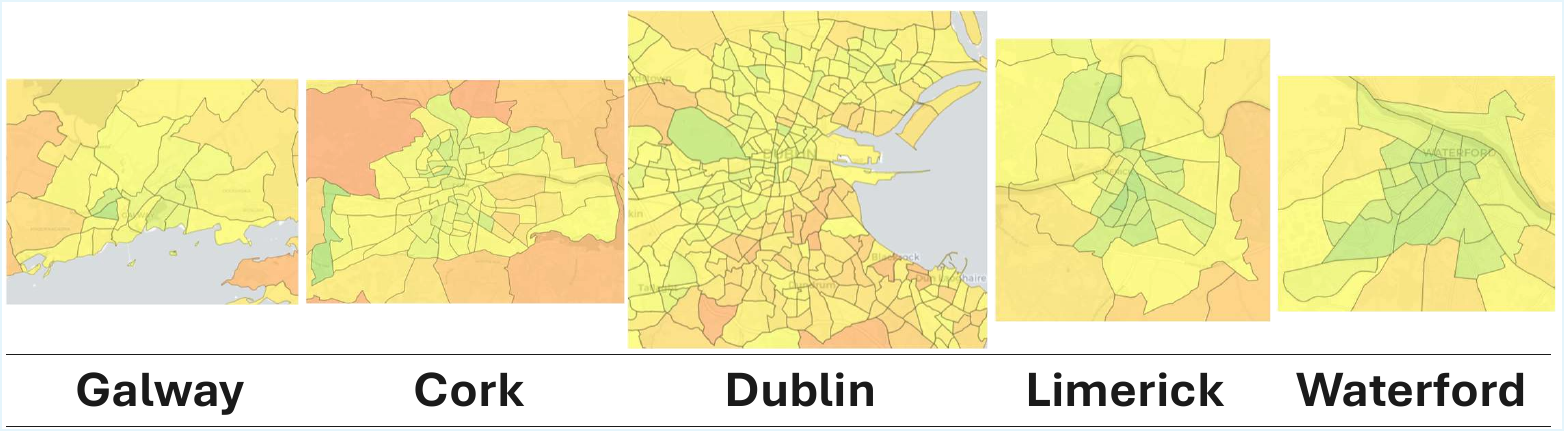}
    \caption{Change in mean SRH for Ireland's 5 cities in the best-case scenario.}
    \label{fig:zoom_on_cities}
\end{figure}

Figure \ref{fig:sideBySideChoropleths} presents the change in mean SRH per ED for each of the three scenarios discussed in previous sections. It is evident in each scenario that a large portion of EDs are projected to see relatively minor changes to the mean SRH in the area. On the other hand, there are some trends which are visible across all three scenarios, such as improvement in SRH for many of the EDs in north-western County Mayo. Another trend evident in all three scenarios is a mild to moderate disimprovement in mean SRH for many of the rural areas in the country. Figure \ref{fig:zoom_on_cities} focuses on Ireland's 5 cities in the best-case scenario and makes it clear that the EDs in these cities are generally projected to experience an increase in their mean SRH. These effects are also evident for the other scenarios, although they are less pronounced.

To explain some of the trends seen in these results, an inspection of individual EDs is required. Although average results from the 5 microsimulation runs for each scenario are useful for insights on the overall outlook for each ED, the output population from a single run must be extracted to trends at the individual level. Therefore, the run which generated the median national population of the 5 runs was chosen as a reference for analysis of single EDs. The ED with the largest average improvement in SRH was Doocharry in County Donegal. In terms of demographic trends, Doocharry's health outlook benefits from the Border region's national high fertility rates. In the reference run, the ED also experiences relatively high levels of net internal migration, gaining a net of 54 people over the course of the microsimulation. As mentioned previously, these internal migrants tend to be younger (and therefore healthier) than the average population. Another contributing factor is Doocharry's relatively old initial mean age of $\approx49$ which decreases down to $\approx44$ and makes it a relatively young ED in 2057.

The ED with the best mean SRH in the country is Cloonteen in Co. Roscommon with a mean SRH of approximately 1.57 (on a scale between 0 and 4). Cloonteen is in the top 200 EDs by youngest mean age of adults, in the top third of most educated EDs (by mean highest level of education achieved) and in the top 50 EDs by proportion of adults who are working. In fact, the ED is projected to possess relatively high proportions of all of the beneficial economic statuses discussed in Section \ref{sec:feature_importances}. Approximately 11\% of the population are projected to be students and the percentage of unemployed people in the area is projected to be just 3\%. 

The ED with the worst average SRH in Ireland is Rathanna in County Carlow. The ED is also projected to have the second largest disimprovement in mean SRH in the country. Analysis of the ED's population quickly reveals why; the ED is only projected to have 19 inhabitants by the end of the reference run and all 19 are aged above 70, with 17 of the 19 being retired. Rathanna's population are also projected to have one of the lowest mean educations nationwide and although approximately $58\%$ of over-70s nationwide are projected to be married, in Rathanna this figure is just 41\%.

One note of caution with regard to the above results is that both the microsimulation and regression models rely on a number of base scenario assumptions which may or may not become a reality. For example, each ED's proportion of Ireland's international immigrants is modelled according to the proportions observed in the year preceding the 2022 Census. 
It is important to recognise that both the regression and microsimulation outputs are conditional on several baseline assumptions. For example, the share of international immigrants to each ED is held constant at levels observed immediately prior to the 2022 Census. Consequently, these estimates should be interpreted as scenario-based forecasts rather than deterministic predictions.
In the SEMIPro paper \citep{me2025irelandMicrosim}, the authors mention that a future extension of the first version of SEMIPro is to add Bayesian modelling of various modules, including international migration. This would facilitate more stochastic sampling of immigrant numbers as well as more rigorous bounds on the uncertainty of the resulting populations. It is also encouraged that users of these models update their assumptions as new data becomes available, and also perform retrospective analyses of how their assumptions compared to the newly observed data. Finally, it is also important to stress that results become more uncertain as the level of granularity increases, as noted in previous microsimulations at the small area level \citep{jia2023norway}. Therefore, the analysis of trends may provide more reliable results than absolute results themselves.


\subsection{Comparison of Emergency Department Distance and SRH}
\label{sec:case_study}

In order to further motivate the study, a case study is presented using the mean SRHs for EDs generated in this study. There is a substantial amount of evidence suggesting that patients who live further away from medical facilities have worse health outcomes than those who live nearby \citep{kelly2016travelTimesVSPatientOutcomes}. Thus, an analysis of the relationship between EDs' mean SRH and their distance from the nearest emergency department was performed. Following on from the precedent set by the CSO in their analysis of each ED's average distance to crucial services in 2019 \citep{cso2019accessToServices}, only emergency departments that are open to adults are considered. The co-ordinates of these emergency departments are retrieved from the Health Service Executive's website. Then, the nearest emergency department to the centroid of each ED is calculated. The CSO's study used the average driving distance for each inhabitant in an ED as their measure of distance. However, much of the data employed in the CSO's study is only available to public authorities and so we consider the ``as the crow flies'' distance to be a suitable placeholder for distance for our purposes. One point to note here is that this analysis is not intending to show a correlation between an area's SRH and their distance from an SRH. There is a link between poor SRH and the need to access medical facilities and there is also a link between travel times to a medical facility and patient outcomes. Therefore, this section is solely intending to highlight some ``interesting'' EDs. For example, a planner may believe that the best location for a new medical facility is close to EDs who are currently very far from such a facility. However, if those far away EDs were largely healthy, and some other closer EDs were much less healthy, it may be wiser to build the new facility closer to these unhealthy EDs.


\begin{figure}[!htb]
    \centering
    \includegraphics[width=\linewidth]{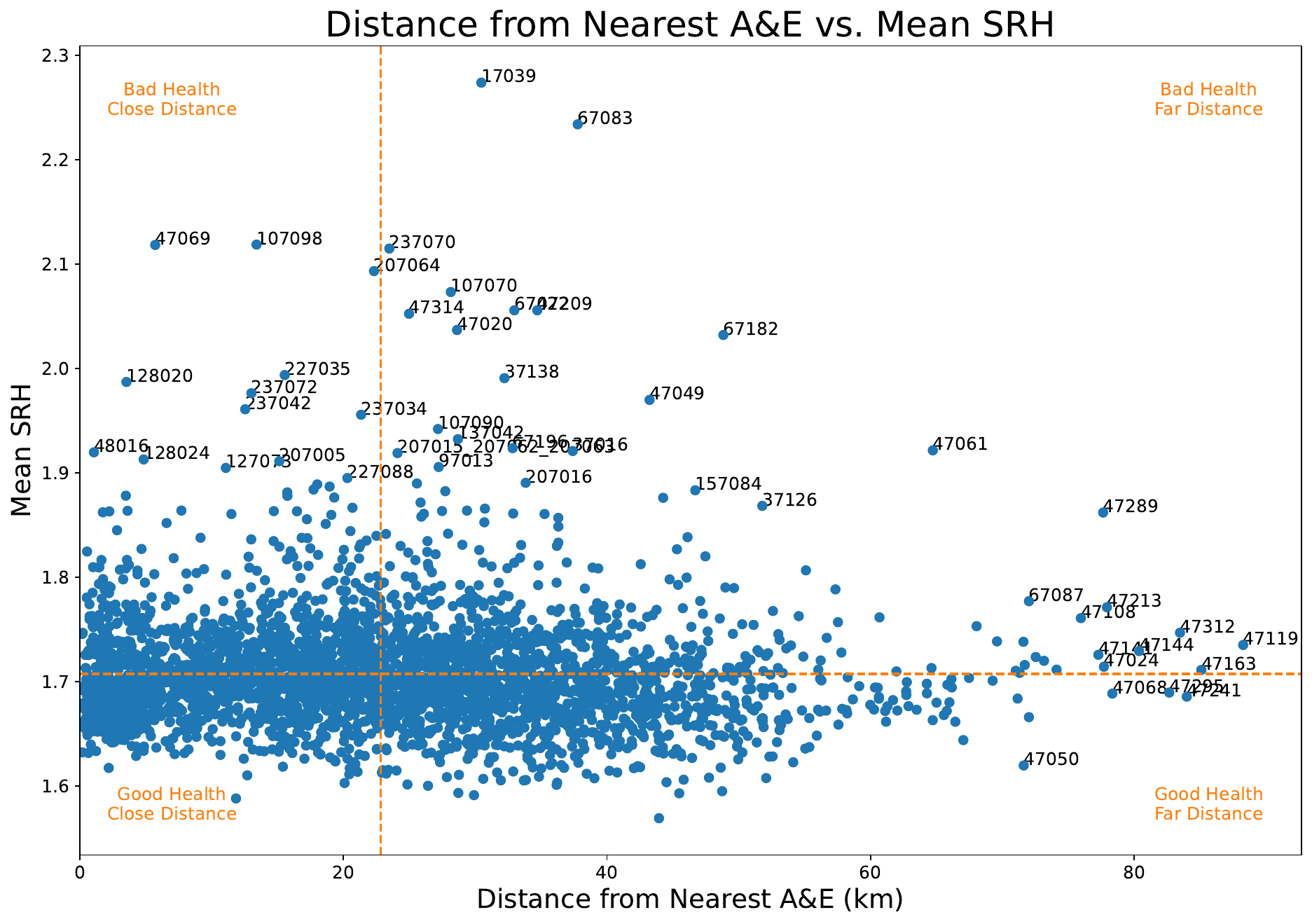}
    \caption{EDs' distance from their nearest emergency department versus their mean SRH. Interesting EDs (far away from the averages) are labelled with their ED\_ID. The dashed horizontal and vertical lines represent the average ED SRH and average distance, respectively.}
    \label{fig:health_vs_distance}
\end{figure}

In Figure \ref{fig:health_vs_distance}, EDs are divided into four quadrants based on their mean SRH and their distance to the nearest emergency department. It is probable that the top-right quadrant, representing worse than average SRH and farther than average distance from the nearest emergency department, will be the quadrant of most interest to the relevant local councils, planning agencies, etc. Towards the far-right end of the graph, a group of ED\_IDs beginning with 47 are visible which corresponds to a group of EDs in the peninsulas of west County Cork. 
An interesting point to note here is that the ED with ID 47049, Bealock, actually resides quite a distance inland of the peninsulas. However, Bealock's mean SRH of approximately $1.97$ is appreciably worse than an ED like Crookhaven, the ED furthest from any emergency department, which has an SRH of approximately $1.74$.
This is a factor which could be taken into consideration when considering the site for any planned new medical facilities, such as an ambulance base. 

\section{Discussion}
\label{sec:discussion}


The M1 microsimulation scenario being considered in the above results is a particularly interesting one, due to the large impact of international migration on population size when compared to natural increase (births minus deaths). International immigrants are generally relatively young, and therefore generally healthy, but international emigrants are equally as young. Therefore, with the net inward migration flows throughout the simulation, the population would be expected to be getting healthier from migration alone. However, this effect has to compete with Ireland's relatively quickly ageing population, as well as its decreasing birth rate. Although children are not included in these SRH regression models, it is possible for children born before 2042 in the microsimulation to be considered in the final SRH predictions (as they will be at least 15 years old). Therefore, the decrease in the proportion of the population in these younger age groups as well as the increase in the proportion of the population in older age groups does add intuition to the overall SRH distributions being slightly worse.

One interesting trend observed in the results for all of the health status distribution scenarios is the relative good health of all 5 of Ireland's cities when compared with rural and suburban areas. Although the migration module in the microsimulation did not determine a migrant's destination based on any of their characteristics, the high immigration flows into these urban areas is likely leading to an influx of healthy individuals. This is because, as shown in the literature, migrants (internal and international) tend to be younger, more likely to be employed and slightly more educated than Ireland's base population \citep{mcginnity2025esriIntegration}. An interesting future direction for this work would be to attempt to model migrants' destinations more specifically, e.g., having young students more likely to move to ``college towns'', and seeing how the results would be affected.

One aspect not considered in this experiment is how the outlook an individual in 2022 holds on their health may compare to an individual with the same characteristics in 2057.
As modern medicine continues to have breakthrough after breakthrough, there is a probability that a 70-year old retiree in 2057 may rate their health better than a 70-year old retiree did in 2022. However, this does not invalidate this model as if the model is incorrect in its prediction of an ED's SRH distribution, it will be incorrect by the same percentage for every other ED also. This means that distributions can still be compared in relative terms, e.g., it would still be interesting to investigate why an ED like Rathanna (Section \ref{sec:small_area_results}) is more unhealthy compared to its neighbouring EDs.

The main limitation of this approach is the uncertainty tied to each of the separate components of the model. There are some small errors in the initial static population and although the 2057 microsimulation population achieved a close match to the size and structure of the DCM-generated population, external validation of the results for the extra socio-economic characteristics was not possible due to a lack of other projections for those characteristics. Furthermore, the predictions of future per-cohort SRH distributions have an inherent level of uncertainty. However, by providing a 90\% confidence level for these distributions, there is a good likelihood that the actual future values of these distributions will end up somewhere in the presented range. Also, by explicitly stating the assumptions for all microsimulation modules, a researcher who may believe an alternate scenario is more likely than the scenarios presented here can easily modify the code to analyse how the results change with their new assumptions.

Another key consideration is the correlation/causation relationship between and individual's characteristics and their SRH. For example, an individual achieving a degree does not suddenly imply that they will rate their health better than before with all else remaining equal. Similarly, becoming unemployed after having a job does not automatically mean that one's health is going to disimprove. It is hoped that it is clear that this paper is not trying to make such claims either. Rather, the proposed method can be used to analyse how the geographical distribution of SRH may look if, in 2057, individuals sharing the same socio-economic characteristics as those in 2022 also share the same outlook on their health.

\section{Conclusion}
\label{sec:conclusion}

This paper has presented a method for predicting the future distribution of SRHs in small areas by combining ordinal regression of health-based microdata and microsimulation. Leveraging of an open-source socio-economic microsimulation along with data linking health to socio-economic characteristics avoids the extremely complex problem of attempting to model how health will change as a result of changes in the direct contributors to healthiness/unhealthiness (obesity, smoking, fitness, etc.).

Validation of the approach is promising, with the mean $R^2$ when comparing actual versus predicted SRH distributions for an ED being approximately $0.9$. Results have been supplied for three different scenarios for future SRHs and in the mean scenario, Ireland is projected to have a slight disimprovement in overall SRH. Although in-depth analysis of why this may occur is not considered to be within the scope of this paper, the aging population is suggested to be the main contributor. 

One avenue for further research from this model would be to apply the same approach to other countries where socio-economic microsimulations are already available such as Canada, the US or Germany, provided that health microdata was also available. In terms of the microsimulation itself, the addition of characteristics describing an individual's ethnicity or nationality could lead to further insights, especially as these factors can have a significant influence on health. This would be a challenging venture, as changes could be required to each module of the microsimulation, provided that data disaggregated by ethnicity was even available. Another possible extension would be a more in-depth case study utilising the future SRH distributions. This could involve incorporating Agent-Based Modelling (ABM) where individuals or even EDs could be treated as agents. A local council could also use the future data to perform an in-depth analysis of their local EDs. 

\newpage


\section*{Declaration of Competing Interests}
The authors declare that they have no known competing financial interests or personal relationships that could have appeared to influence the work reported in this paper.

\appendix
\section*{Appendices}
\renewcommand{\thesection}{A.\arabic{section}}

\section{Repository Link}
\label{sec:link}
All code for the method can be found at \url{https://github.com/SCC-git/srh_ordinal_regression}. Note that Healthy Ireland microdata must be applied for through the ISSDA (\url{issda.ucd.ie}).

\section{Characteristic Values}

\label{sec:characteristicValuesTable}
    \begin{longtable}{|C{0.25\linewidth}|C{0.475\linewidth}|C{0.15\linewidth}|}
    \hline
    \textbf{Characteristic} & \textbf{Possible Values} & \textbf{Value in Source Code}\\
    \hline
    \multirow{1}{\linewidth}{Age} & 0-105 & Relevant Integer\\
    \hline
    \multirow{2}{\linewidth}{Sex} & Female & F \\
    & Male & M \\
    \hline
    \multirow{4}{\linewidth}{Marital Status} & Married & MAR \\
    & Single & SGL \\
    & Separated & SEP \\
    & Widowed & WID \\
    \hline
    \multirow{4}{\linewidth}{Citizenship} & Ireland & IE \\
    & UK & UK \\
    & EU27 excluding Ireland & EU \\
    & Rest of World & RW \\
    \hline
    \multirow{1}{\linewidth}{Moved to Ireland in last year} & True or False & Relevant Boolean \\
    \hline
    \multirow{8}{\linewidth}{Highest Level of Education Attained} & No Formal Education & NF \\
    & Primary Education & P \\
    & Lower Secondary & LS \\
    & Upper Secondary & US \\
    & Post-Leaving Certificate & PLC \\
    & Higher Certificate & HC \\
    & Undergraduate Degree & DEG \\
    & Postgraduate Degree & PD \\
    & Doctorate & D \\
    \hline
    \multirow{8}{\linewidth}{Primary Economic Status} & At Work & W \\
    & Student & S \\
    & Looking After Home/Family & LAHF \\
    & Retired & R \\
    & Unable to Work Due to Permanent Sickness or Disability & UTWSD \\
    & Other & OTH \\
    & Unemployed & UNE \\
    & Not Applicable (for children) & NA \\
    \hline
    \caption{The characteristics used to describe individuals along with their values in the source code}
    \label{tab:characteristicsAppendix}
    \end{longtable}
    


\end{document}